# Modular Resource Centric Learning for Workflow Performance Prediction

Alok Singh, Mai Nguyen, Shweta Purawat, Daniel Crawl, Ilkay Altintas
San Diego Supercomputer Center
University of California, San Diego
La Jolla, CA, U.S.A.
a1singh@eng.ucsd.edu, {mhnguyen, shpurawat, crawl, altintas}@sdsc.edu

*Abstract*— Workflows provide an expressive programming model for fine-grained control of large-scale applications in distributed computing environments. Accurate estimates of complex workflow execution metrics on large-scale machines have several key advantages. The performance of scheduling algorithms that rely on estimates of execution metrics degrades when the accuracy of predicted execution metrics decreases. This *in-progress paper* presents a technique being developed to improve the accuracy of predicted performance metrics of large-scale workflows on distributed platforms. The central idea of this work is to *train resource-centric machine learning agents* to capture complex relationships between a set of program instructions and their performance metrics when executed on a specific resource. This resource-centric view of a workflow exploits the fact that predicting execution times of sub-modules of a workflow requires monitoring and modeling of a few dynamic and static features. We transform the input workflow that is essentially a directed acyclic graph of actions into a Physical Resource Execution Plan (PREP). This transformation enables us to model an arbitrarily complex workflow as a set of simpler programs running on physical nodes. We delegate a machine learning model to capture performance metrics for each resource type when it executes different program instructions under varying degrees of resource contention. Our algorithm takes the prediction metrics from each resource agent and composes the overall workflow performance metrics by utilizing the structure of the corresponding Physical Resource Execution Plan.

## I. INTRODUCTION

As high-performance computing applications evolve to embrace exascale computing in the future, declarative directives provide a preferred model of programming [2]. Workflows provide an expressive programming model for fine-grained control of such large-scale applications in distributed computing environments. A workflow declares a set of computing tasks over the data. It leaves the task of implementation to the execution engine, which can utilize resource-specific knowledge to optimize both application execution and resource utilization. This decoupling of implementation from application development simplifies the application development process [26].

The workflow programming model leaves the task of optimal scheduling and resource utilization to the underlying execution engine. The field of scheduling and resource provisioning on large-scale systems has been an active area of research. Several scheduling algorithms [3, 4, 5, 6] depend on accurate estimates of runtime information and data flow. The performance of scheduling algorithms that rely on estimates of execution metrics degrades when the accuracy of predicted execution metrics decreases [1]. We present a methodology that utilizes machine learning (ML) techniques to predict resource usage and execution times of complex workflows on distributed platforms. Our proposed methodology learns from past workflow executions to train resource-specific models. This technique scales well with diverse workflows since it focuses on learning by characterizing the sub-components of a workflow and characterization of computing resources. Our prediction engine trains a model for each resource type to capture the resource utilization patterns of different program executions on that resource, i.e., how the underlying resource behaves when sub-modules of a workflow execute on it. Each new workflow execution helps the resource level learning models to improve their understanding of resource utilization behavior of given instruction sets.

We propose a framework for performance prediction of arbitrarily nested workflows that run on distributed platforms. Our technique views a workflow as a collection of sub-modules running on specific resources and performs localized learning for each resource site. It utilizes instruction set characterization, machine configuration and system workload information to predict overall workflow performance metrics. Specifically, we aim to make following contributions through this research:

1. Identify essential characteristics of hardware resources, program instructions and system load to make accurate predictions of performance metrics of workflow execution instances.
2. Develop a modular ML-based model that trains resource-specific agents to learn the behavior of modular building blocks of a large-scale workflow.
3. Demonstrate that this modular technique scales to large workflows involving arbitrary levels of nested tasks and complex dataflow patterns.
4. Empirically show that resource-node level predictors deliver a scalable solution for wide range of workflows from a relatively small training sample.

The paper is organized as follows: Section II presents an overview of the proposed approach, including the performance metrics, the data collection process, ML models, and features for characterizing a workflow. Section III details a representative workflow for experimental validation of the hypothesis. Section IV overviews related work in the area of application and workflow performance prediction. Section V summarizes the planned future work.

## II. APPROACH

We propose a performance prediction approach that utilizes resource-specific learning agents. A workflow is an abstract representation of computational tasks through which the input data passes [26]. It shows how to transform and transport the input data and the order in which computational tasks perform this transformation. Given a workflow, our execution engine constructs a *Physical Resource Execution Plan* (PREP), which is a mapping of subtasks of the workflow to specific compute nodes in a cluster. The PREP defines precise hardware execution site(s) for each task and the order in which the data will flow between them. [26] presents one such tool that performs this mapping. In the current work, we focus only on workflows that have synchronous data flows between two sequential computational tasks. In other words, when two tasks are connected sequentially, the successor starts after the predecessor ends computation.

The main contribution of our work is to deploy a ML model (called agent) for each physical node to capture its behavior when the resource is executing a sub-module of the workflow. Since we characterize the sub-modules in an application-independent way that captures its instruction level features, the resource-centric agent can predict performance for newer sub-modules.

The physical resource execution plan decouples the task of data movement from instruction execution. This decoupling enables us to efficiently capture the characteristics of somewhat disjoint tasks (data movement over a network vs. computation on a node) in separate ML models.

During the *profiling phase*, we run workflows on multiple distributed platforms, and profile the performance data on a per node basis. The profile data captures each node's static characteristics, program instructions' execution characteristics and the environment's dynamic characteristics as input parameters to the learning agents. During the training phase, the agents are trained to capture the relationship between these characteristics and the output parameters.

During the *prediction phase*, the prediction engine takes a PREP as input and calls each of the resource-specific model to perform performance execution site-specific predictions. The post-processing step uses these predictions and the structure of PREP to generate an overall performance prediction for the entire workflow. Nested workflow predictions can be processed recursively in a bottom-up manner.

Each *resource-centric model* learns from a variety of workflows that execute sub-modules on that node. Since the sub-modules are characterized using application-independent features, each *compute node model* is trained to capture resource behavior of a set of machine-level instructions. Even though two workflows might have different applications and may accomplish different tasks, they might have overlapping sets of machine-level tasks, e.g., each workflow might have sub-tasks that perform complex floating-point operations (FPOs) in different areas of the execution plan. The resource model trained by one workflow can be used to predict a sub-module's behavior that performs similar FPOs for a different workflow. Resource-centric learning and application-independent characterization of instructions can enable cross workflow prediction and can scale to predict the performance of diverse workflows from limited training examples.

### A. Performance metrics

In the execution of a workflow, several different types of resources are used, each with different characteristics and behavior. Thus, performance metrics based on different resource types must be predicted to provide accurate information for effective scheduling of resources. The main categories of resources are processor (CPU and GPU), memory, I/O, and network. To characterize usage of these resource types, the following metrics are predicted: execution time, memory usage (peak and average), I/O access time, and network transfer time.

In our modular approach to performance prediction, these metrics are predicted for each building block, then aggregated to form predictions for the entire workflow.

### B. Workflow Execution Characterization

In order to predict the performance metrics listed in the previous section, a set of parameters is required to accurately capture the different resources used during execution of a workflow. The utilization of each resource depends on factors that are *static* (to characterize the resource) and *dynamic* (to capture the resource's behavior in the run-time environment). Additionally, there are parameters influencing resource utilization by the *application* itself, e.g., input dataset size significantly affects execution time and memory usage. Thus, we specify three categories of parameters to characterize resource utilization: static, dynamic, and application-specific. Examples of parameters in these categories include process speed and memory size (*static*); number of jobs running and queue waiting time (*dynamic*); and number of floating-point operations and branching factor (*application-specific*). These parameters are used as input to the models (see: Section D) to predict the above-mentioned performance metrics.

### C. Data Collection

As described above, we will collect information about the computational environment, applications, and runtime performance. Static information about the computational hardware and applications can be collected from sources such as the Linux *proc* file system (http://www.tldp.org/LDP/Linux-Filesystem-Hierarchy/html/proc.html). The *proc* file system contains directories and files that provide an interface to various kernel data structures, e.g., the number of cores, CPUs, CPU speed, and cache sizes from /proc/cpuinfo, and the amount of memory from /proc/meminfo.

Information about dynamic parameters includes memory usage, cache hit rate, and execution time. These parameters should be captured with minimal overhead since we do not want the overhead to influence the ML model and performance predictions. Many of these parameters can also be collected from the *proc* file system: each running process has a directory /proc/<pid>/ that contains files providing information about the process. For example, /proc/<pid>/status includes the current virtual memory size, the maximum virtual memory size, the current resident set size, and the number of threads. The files in /proc/<pid> can periodically be read to obtain snapshots of the application's performance without incurring a large overhead.

The Kepler [7] provenance framework also collects runtime information such as the overall workflow execution time and the amount of execution time for each actor. This information, along with the static and dynamic profiling parameters will be stored in a centralized repository.

## D. Machine Learning Models

We use a ML approach to construct a model to predict performance metrics such as execution time. In our modular approach to performance prediction, predictive models are created for metrics of each sub-module of a workflow instead of the entire workflow.

Resource utilization prediction will be achieved by presenting a model with historical data with input attributes characterizing a sub-workflow executing on a computing platform, along with the actual value of a target performance metric. A ML algorithm is used to adjust the parameters of the model such that the mapping between input data and the target is learned by minimizing a loss function. In our approach, parameters that characterize a sub-workflow and the computing environment, as described in Sec B, are used as input data to a model, and the model is trained to predict a performance metric such as execution time. Once trained, the model can be used to predict resource utilization for a similar but previously unseen sub-workflow, i.e., one that was not used to train the model. The data to train and test our predictive models comes from our repository of historical data as described in Section C. The metrics of interest, namely execution time, memory usage, disk access time, and network transfer, are all continuous-valued entities; therefore, we use regression techniques such as neural networks and random forest to construct our prediction models. Metrics for each resource type can be predicted separately since different resource types have different characteristics.

We envision generic models that predict performance for a general class of hardware for a resource type, as well as specialized models that are specific to a particular computing configuration. For example, Gordon and Comet at SDSC, two XSEDE (See: xsede.org) computing resources, are used to create models specifically for these clusters. The Gordon model can be used to predict execution time of a sub-workflow running on Gordon, and similarly for workflow executions on Comet. We also plan to combine sub-workflow execution data for both Gordon and Comet to create a model for predicting performance metrics for a generic high-performance computing environment. Generic models will provide less accurate predictions than specialized models, but are useful since they are applicable to scenarios in which we do not have data for a specific computing environment.

## E. Predicted Temporal Graph

The models described in the previous section are used to predict utilization of a resource for each sub-module in a workflow. Since a workflow can be viewed as an execution plan of sub-modules over a set of physical resources, individual sub-module predictions can be aggregated together to form performance predictions for the entire workflow. A choice of resources for a resource type will present as alternate paths through the Physical Resource Execution Plan, and predictions for the different alternatives can be compared to determine the optimal choice for the workflow.

By decomposing a workflow into smaller modules, data collection and model training are simplified. In addition, modules can be assembled in different way, allowing for complex and new workflows to be represented. Thus, our modular approach allows performance of complex workflows to be predicted efficiently. It also makes performance prediction of new workflows possible.

## III. REPRESENTATIVE USECASE: MTGA WORKFLOW

To demonstrate the application of the proposed modular technique to a large workflow, we are using the bioKepler [27] implementation of Microbiome Taxonomy and Gene Abundance Workflow (MTGA) as a representative use case. The workflow performs comparative gut microbiome analysis across human gut microbiome of patients with autoimmune diseases and healthy subjects [8]. The workflow comprises of analysis stages with varied level of nesting and parallelization ranging from single core to several cores on multiple nodes. Additionally, all the stages deal with large amounts of input data and produce similar sized output data sets.

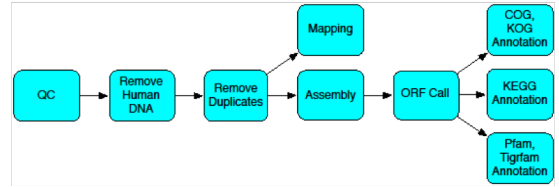

Fig. 1. Read and assembly-based MTGA workflow [8]

### A. Computational Characteristics of Sub-Modules

MTGA (Fig. 1) is both data- and compute- intensive. Table I shows the compute requirements of major steps of the workflow. Each stage uses a different software tool with distinct computing demands.

TABLE I. STEPS OF COMPUTATIONAL TOOLS IN MTGA WORKFLOW

| Analysis | Tool | Data (Input Size, Ref DB Size) | CPU (Cores, Nodes) | CPU usage (core-hours) | Peak Memory Usage per node |
|---|---|---|---|---|---|
| Quality control | QC script | 43GB | 1, 1 | 38 | ~10 MB |
| Remove Human DNA | Bowtie | 9GB, 6GB | 16, 1 | 2 | ~10GB |
| Remove Duplicates | CD-HIT-DUP | 7GB | 16, 1 | 670 | 256GB-512GB |
| Mapping | FR-HIT | 3.4GB, 16GB | 16, 32 | 4784 | ~210GB |
| Assembly | Velvet | 7GB | 16, 1 | 700 | 256GB-512GB |
| ORF call | Metagene | 200MB | 16, 1 | 5 | 0.5GB |
| Annotation (Pfam) | HMMER 3 | 90MB, 1GB | 16, 8 | 355 | ~5GB |
| Annotation (KEGG) | BLASTP | 90MB, 6GB | 16, 16 | 11960 | ~10GB-30GB |

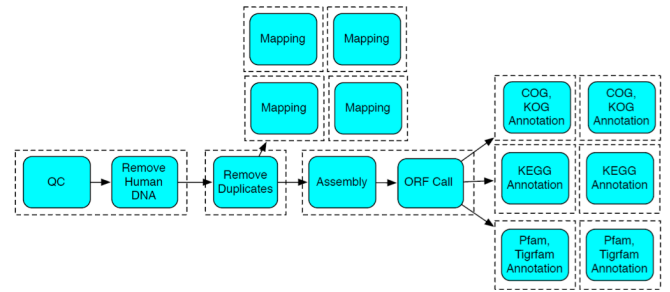

Fig. 2. Example Physical Resource Execution Plan for MTGA workflow

The SDSC Gordon supercomputer's features such as large RAM settings, ultra-fast Oasis file system are utilized to execute and profile the MTGA workflow. To incorporate the impact of input data size on performance metrics, we are profiling the workflow execution traces for a range of input sequence datasets. Fig. 2 shows a sample physical mapping of the logical workflow to resource nodes. Each dotted box represents a compute resource node.

## IV. RELATED WORK

The task of predicting execution times and resource requirements of complex applications has been an active area of research and several works focused on the use of ML techniques for scientific workflow performance prediction.

[9] utilizes similarity functions to find nearest neighbors dynamically and then apply induction methods to predict execution times. The approach uses a single global induction model that takes input features to predict the output, deploying three induction models: k-NN, k-Weighted Average, Locally Weighted Linear Regression. [10] presents a level based prediction model that centers its prediction on the structure of the input workflow. It organizes the workflow into levels, where tasks at the same level are independent of each other. A level-based estimation model is deployed to make predictions. [11] compares several ML modeling techniques for bioinformatics workflow prediction (BLAST, RAxML). The workflows are characterized using application level features, advocating usage of a vast number of attributes and leaving the task of making the best use of them to the algorithm. In [12] a workflow is divided into blocks through which the data flows. The modeling technique utilizes block parameters and the data volume traveling among them. It deploys a linear model for prediction. However, the experiments are performed without using the machine characteristics.

[13] uses similarity templates to predict execution times of scientific workflows. [14] studies the influence of input features on the predicted values by using a ML algorithm called C4.5 decision tree builder. The variability due to cross-platform execution is not observed and all experiments on a single idle machine. The NIMO [15] system generates resource assignments for scientific workflows on large-scale networks, and characterizes the application behavior, data, and the underlying hardware as feature vectors. However, the predictor functions in NIMO are limited by the assumption that the output value is a linear combination of pre-determined transformations of the input characteristics. Forecasting models based on Support Vector Machines, Neural Networks, and Linear Regression are presented in [16], in which the focus is on predicting the performance of the TPC-W benchmark by deploying a web server on a virtual machine and a database server on a separate machine. This work develops a single training model for predicting the web application's resource needs. In contrast, we present a modeling technique that trains a large number of agents to learn behavior of each resource type when they run a given set of instructions. [17] presents ANN and SVM based modeling technique to predict application behavior in a virtual machine environment as a function of VM configuration parameters. They demonstrate that sub-modeling presents improved results compared to a single global prediction model.

Several other works use various modeling techniques to predict the performance of applications ranging from stand-alone programs and web services benchmarks to applications running in virtualized environment.

In [18], a combination of power law model, queuing models and request mix models to perform the impact on response times due to architectural and workload changes. Sub-models are calibrated to predict an attribute of the overall performance metric from system parameters and composed to predict application performance under a new situation. However, this proposed composition method uses expert level knowledge of the structure of Internet services.

[19] presents an abstract machine model based on Fortran. Execution times of each Fortran program is expressed as the linear sum of execution times of individual Fortran abstraction operations (called AbOps) multiplied by its frequency of execution. The paper characterizes a machine by using a set of abstract operations representative of the language constructs found in Fortran. It performs static program analysis by finding the frequency of these operations in the source code. The dynamic analysis is performed by measuring the number of times each line of source code is executed. The execution times prediction is performed by combining machine and program characterizations. [20] presents a cost model to estimate required environment configurations for scheduling workflows in the cloud environment, by optimizing the execution time and the monetary cost of execution. [21] presents an online method to automatically characterize the resource needs of tasks in a workflow based on profiled information. The method finds a correlation between the input data size and resources needed such as CPU usage, Memory consumption and execution time to predict task needs of a workflow. [22] proposes a smoothing method that performs prediction of execution times in Grid environments. The proposed method handles the sudden peak changes and level switches present in Grid computing. In [23], a probabilistic model to predict workflow makespan is given. It captures the variability of the grid environment using a random latency variable. [24] offers an approach that obtains runtime predictions of online tasks from several strategies and choose the best one based on a proposed evaluation criterion. The method does not distinguish between the local and the remote tasks. [25] presents a pattern based time-series prediction approach that makes a distinction between long duration and short duration workflow activities.

Our work looks at the problem of prediction workflow performance from a resource-centric perspective. We attempt to reduce the complexity of characterizing a large-scale workflow execution on a distributed platform by delegating the task of prediction to several resource specific agents. Each of these agents learns the non-linear relationship of executing a sub-module of a workflow on a given hardware. In this sense, each hardware node's agent specializes in predicting performance metrics of executing a smaller set of instructions on that node. This strategy reduces the number of interactions each model has to learn and thus improves prediction accuracy. Secondly, we characterize each sub-module based on instruction level features. Hence, our technique is likely to scale well for predicting the performance of new workflows.

## V. CONCLUSIONS AND FUTURE WORK

In this paper, we present a resource-centric ML approach for prediction of large-scale workflow performance metrics. A workflow defines a set of computational tasks that execute in a particular order on the input data. An execution engine decides the optimal execution strategy to perform the declared computations on a set of available resource nodes. We call this strategy a Physical Resource Execution Plan (PREP). The PREP contains information about the resource nodes and the workflow sub-modules that execute on those nodes.

The proposed technique utilizes ML models to capture the behavior of a sub-module's instruction set when it runs on a particular resource node. We characterize each sub-module in an application-independent way to capture the computational load that this program represents. We are investigating different techniques to perform this characterization in an efficient way. Other features that will influence the output of the trained learning model are hardware specifications, background traffic (resource contention) and input data size.

Resource-centric learning reduces the problem of capturing the behavior of a large-scale workflow on a distributed platform to many smaller learning tasks. This modular approach is likely to scale well across workflows because we characterize each sub-module's machine-level footprint and train sub-models rather than training one large model to learn everything about a workflow's complex execution pattern. We plan to implement and extend the outlined modeling steps and use the MTGA workflow [8] to validate our approach.

ACKNOWLEDGMENTS

The authors thank our colleagues in the IPPD project for their collaboration and Arvind Rao for his help on profiling of workflows. This work is supported by NSF DBI 1062565 and 1331615, NIH P41 GM103426 for NBCR and R25 GM114821 for BBDTC, and DOE DE-SC0012630 for IPPD.